\documentclass[aps,prb,twocolumn]{revtex4}
\usepackage{graphics,amsmath}
\begin{document}
\title{Dynamical mean field theory of small polaron transport}

\author{S. Fratini} 
\affiliation{Laboratoire d'Etudes des Propri\'et\'es Electroniques des Solides, CNRS - BP166 - 25, Avenue des Martyrs, F-38042 Grenoble Cedex 9} 
\author{S. Ciuchi}  
\affiliation{Istituto Nazionale di Fisica della Materia and  
Dipartimento di Fisica\\ 
Universit\`a dell'Aquila, 
via Vetoio, I-67010 Coppito-L'Aquila, Italy} 

\date{\today}
 
\begin{abstract}
We present a unified view of the transport properties of
small-polarons in the Holstein model at low carrier densities, 
based on the Dynamical Mean Field Theory. 
The nonperturbative nature of the approach allows us to study the
crossover from classical activated motion at high temperatures to
coherent motion at low temperatures. 
Large quantitative discrepancies from the standard polaronic
formulae are found. The scaling properties of the resistivity are analysed, and
a simple interpolation formula is proposed in the nonadiabatic
regime. 
\end{abstract}
\maketitle

In the common wisdom, polaronic transport in solids is synonymous of activated
conductivity. When electrons  are strongly coupled to bosonic degrees
of freedom (phonons, excitons, etc.), 
self-trapped states are formed. If the size of the polaron is 
comparable with the lattice spacing,  the motion is then dominated by hopping
processes: the particle has to overcome a potential barrier $\Delta$
and loses its quantum coherence at each hop, giving rise to an
activated law of the form:
\begin{equation}
  \label{eq:activated}
  \rho=\rho_0 e^{\Delta/k_BT} 
\end{equation}
where the prefactor $\rho_0$ is weakly temperature dependent. Contrary
to what happens in semiconductors, the activated behaviour here is not
related to the number of thermally excited carriers, but rather to
the mobility of the individual carriers.

The Holstein model\cite{Holstein,Tjablikov} 
\begin{equation*}
  \label{eq:Holstein-model}
  H= - t \sum_{i,j} (
 c^\dagger_{i} c_{j}  + c^\dagger_{j} c_{i} )  + \hbar \omega_0\sum_{i}
 a^\dagger _{i} a_{i}  -g \sum_{i}   c^\dagger_{i} c_{i}
 (a^\dagger _{i}+a_{i}) 
\end{equation*}
was introduced in the late fifties to study
such a behaviour, as was being measured in some transition metal 
oxides. In this model, tight-binding electrons 
interact locally with  molecular deformations, whose natural vibration 
frequency is $\omega_0$ ($t$ is the hopping
parameter, $g$ the electron-phonon coupling constant).
Although this is a rather crude idealization of a real solid, the
Holstein model captures 
the essential physical phenomena involved in small-polaron transport.
The situation is in fact more complex than indicated by the simple
formula (\ref{eq:activated}), and 
is summarized in several
reviews\cite{Appel}. Three regimes of temperature can
be identified.

At low temperatures,
the polarons behave as heavy particles in a band of
renormalized width $W$, and are weakly scattered
by phonons. 
For small enough $W$, all the states in the band are
equally populated, leading to a ``metallic-like'' resistivity 
\cite{FrohlichSewell,LF,Appel}
$  \rho \propto (T/W) \exp (-\hbar\omega_0/k_BT)$.
The exponential law comes from the thermal occupation of the optical
phonons, which are assumed to be the main source of scattering.
Upon increasing the temperature, the mean free path is rapidly reduced
until the picture of coherent motion breaks down, 
typically around some fraction of $\hbar\omega_0/k_B$. 
Hopping motion then becomes more favorable, leading to an activated
behavior of the form  (\ref{eq:activated}).
The crossover from coherent to hopping motion is thus characterised by
a maximum in the resistivity.
Eventually, at temperatures higher than the activation energy, the
polaron states are thermally dissociated and the residual electrons
are strongly scattered by thermal phonons. In this case, the 
equipartition principle leads to \cite{Adler} $\rho\sim T^{3/2}$.  


On the experimental side, the largest amount of work has been devoted
to the activated regime, which is often observed around room temperature,
and resistivities of the form
(\ref{eq:activated}) have been measured in a variety of narrow-band solids. 
However, strong deviations from  pure Arrhenius behaviour
 are often reported\cite{deviations}, possibly indicating the onset of the
coherent transport regime. In some cases, the low temperature exponential law
described above has also been identified  unambiguously 
\cite{coherent}.

The main purpose of this work is to shed some light on the
crossover from activated to coherent transport, 
for which a reliable theoretical description is still lacking. 
To do this, we calculate  the
resistivity of  Holstein small-polarons in the
framework of  the Dynamical Mean Field Theory (DMFT).
This approximation is suitable whenever the
physics is ruled by local phenomena, as is the case in the present
problem, where it allows to take into account the quantum nature of the 
phonons ($\omega_0\neq 0 $) and the 
finite bandwidth effects ($t\neq 0$) on the same footing. 
Since the theory does not require any ``small parameter'', it is able to
go beyond the traditional approaches usually applied to the problem and
gives reliable results in the regime $k_B T\sim \hbar \omega_0$ of
interest here. Moreover, it yields a unified view of the different regimes
of polaronic transport, acting  as a testing ground of
the validity of previous approaches


The DMFT solution of the Holstein model for a single polaron
was presented in ref.
\cite{depolaroni}, where an analytical expression for the 
spectral function $A_\epsilon(\nu)$ was given  
in terms of a continued fraction. 
The polaron formation
{\it at zero temperature} can be described by
introducing two independent dimensionless parameters. 
The first is the adiabaticity
ratio $\gamma=\omega_0/D$ ($D$ is the unrenormalized half bandwidth) according to which 
an adiabatic ($\gamma\ll 1$) and nonadiabatic regime ($\gamma\gg1$) can be 
defined.
The mechanism of polaron formation is fundamentally 
different in the two regimes, leading to different definitions of the
dimensionless electron-phonon coupling. 
Being $E_P=g^2/\omega_0$ the energy of a polaron on a single
lattice site, a well defined polaronic state is formed for large 
$\lambda=E_P/D$ in the adiabatic case, and for 
large $\alpha^2=E_P/\omega_0$ in the non
adiabatic case\cite{depolaroni,xover}.
%

The corresponding transport properties can be calculated through the appropriate 
Kubo formula, which relates them 
to the current-current correlation function of the system
at equilibrium. In DMFT, due to the absence of vertex
corrections,\cite{RMP} the latter  
is fully determined by the  spectral function 
$A_\epsilon(\nu)$, which is known \textit{exactly} in the limit of
vanishing density (single polaron problem)\cite{depolaroni}.  
The resistivity at low (but finite) density  can then be derived 
through an expansion in the fugacity\cite{BR}, yielding
\begin{equation}
  \label{eq:rho}
  \rho(T)=\frac{k_B T}{x \zeta \pi} 
\frac
{  \int d\epsilon N_\epsilon
 \int d\nu  e^{-\nu/T}  A_\epsilon(\nu)
}{\int d\epsilon N_\epsilon \phi_\epsilon
 \int d\nu  e^{-\nu/T}
          [A_\epsilon(\nu)]^2}
\end{equation}
In the above formula, 
the denominator is the current-current correlation function, and
the numerator is proportional  the  carrier concentration $x$, which
was explicitely taken out (the exponential weights in the integrals 
reflect the Boltzmann nature of the carriers). 
$N_\epsilon$ and $\phi_\epsilon$ are  respectively the density of states and
the current vertex of the periodic lattice.
\footnote{We specialize here to a Bethe lattice with
  semicircular density of states, which is representative of
  three-dimensional systems,  and use the prescription given in 
[A. Chattopadhyay  \textit{et al.},  Phys. Rev. B {\bf 61}, 10738
(2000)]. However, since the Kubo formula involves averages over the whole
density of states, the results should not depend dramatically on the
shape of  $N_\epsilon$.} 
The constant 
$\zeta=e^2a^2/\hbar v$ has the dimensions of resistivity, $a$
being the lattice spacing, $v$ the volume of the unit cell.
In the following, we shall implicitely report the
results for the dimensionless product $\rho x \zeta$, which is inversely
proportional to the drift mobility, and assume $\hbar=k_B=1$.


%
\begin{figure}[htbp]
\centerline{\resizebox{8cm}{!}{\includegraphics{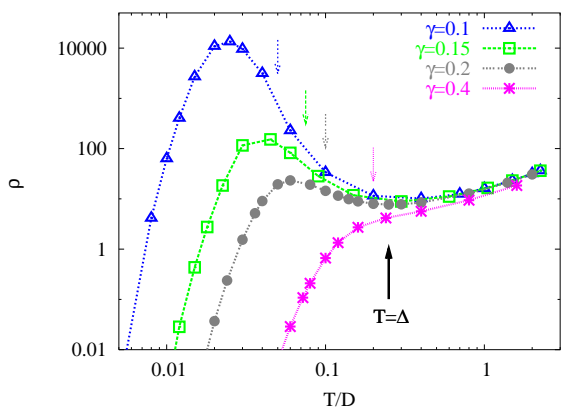}}}
\caption[]{Resistivity 
vs. temperature at $\lambda=1.5$, for different values of
the adiabaticity parameter. 
Short arrows mark the temperature $T=\omega_0/2$  
below which phonon quantum effects acquire importance. The long upward arrow
is the gap $T=\Delta$. }
\label{fig:adiab-fixedlambda}
\end{figure}
The results for the resistivity are
illustrated in Fig. \ref{fig:adiab-fixedlambda} for fixed $\lambda=1.5$ 
in the adiabatic case. 
The three regimes discussed in the
introduction can be clearly identified 
--- the resistivity first rises
exponentially (coherent regime),  then decreases exponentially
(activated regime), and eventually 
increases again as a power law (residual scattering
regime). This is true for all the data sets except at $\gamma=0.4$, 
where the polaron formation has shifted to 
higher values of $\lambda$, as expected when moving away from the
adiabatic limit.  
\cite{depolaroni,xover} 

Let us focus on the activated regime,  
$\omega_0\lesssim T\lesssim\Delta$,  
where the polaron transport is dominated by incoherent hopping processes.
In the adiabatic limit $\gamma\to 0$, the problem
is generally studied within   a simplified ``two-site'' molecular
model\cite{Holstein,LFJETP68}, where the (classical)  
lattice degrees of freedom are seen to  move ``adiabatically'' 
in the double-well energy curve 
determined by the electronic (bonding) state. 
At each jump,  the system has to overcome a barrier $\Delta=E_P/2-t$,
leading to an Arrhenius type behavior\cite{Appel} 
\begin{equation}
  \label{eq:adiab-act}
  \rho=2(T/\omega_0) \exp[\Delta/T]
\end{equation}
This semi-classical description holds provided that the transitions 
to higher (antibonding) electronic states can be neglected, which
corresponds to 
\cite{Holstein,LF63}  
\begin{equation}
  \label{eq:condition}
   \eta_2\equiv\frac{D^2}{\omega_0 \sqrt{2E_P T}}=
\left\lbrack 2 \lambda \gamma^3  (T/\omega_0) \right\rbrack^{-1/2}\gg 1 
\end{equation}
Note that this does not coincide with the usual adiabaticity
condition $\gamma\ll 1$ relevant for polaron formation. 

%
\begin{figure}[htbp]
\centerline{\resizebox{8cm}{!}{\includegraphics{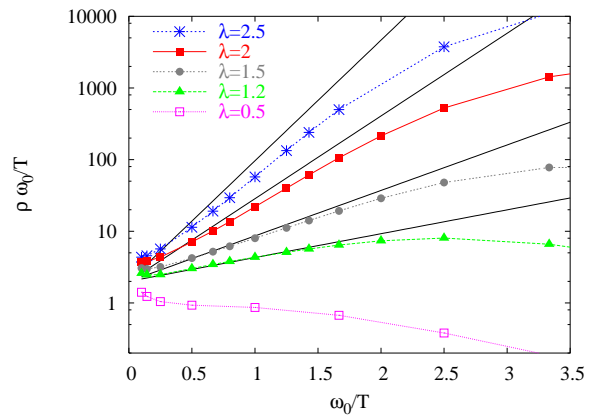}}}
\caption[]{Arrhenius plots of resistivity divided by $T/\omega_0$ for fixed 
$\gamma=0.2$, at various $\lambda$.  The DMFT data are
compared with the semi-classical formula (\ref{eq:adiab-act}) with $
\Delta=[E_P(\lambda)-D]/2$ 
(straight lines) --- see text. }
\label{fig:adiab-fixedgamma}
\end{figure}
%
To illustrate the accuracy of the semi-classical prediction,  we
show in Fig. \ref{fig:adiab-fixedgamma} Arrhenius plots of the
resistivity at 
fixed $\gamma=0.2$, varying the coupling strength $\lambda$.
First of all, our results indicate that the correct generalization 
of the ``two-site'' result  (\ref{eq:adiab-act}) to infinite lattices 
is obtained by letting  $ \Delta=[E_P(\lambda)-D]/2$, where 
$E_P(\lambda)=D\lambda  +D/(8\lambda)+\cdots$  is the adiabatic polaron
energy, calculated e.g. in ref. \cite{depolaroni}  [see Fig. 2, full
lines --- the slight discrepancy at the highest values of $\lambda$
is related to the breakdown of the condition (\ref{eq:condition})]. 
This suggests that the activated behavior arises from the thermal 
promotion from the ground state to the electron continuum, which 
differs from the Landau-Zener mechanism involved in the two-site model.
In particular, the reduction of the activation gap by finite bandwidth
effects is much stronger in the present case.


When the temperature is lowered below $T\approx \omega_0$, 
the quantum nature of the phonons can no longer be neglected. The
lattice zero point fluctuations are then expected to induce
delocalization of the 
trapped carriers, resulting in an  enhancement of the polaron
mobility. This phenomenon is a precursor of the coherent regime,  and 
has a sizeable influence on the
transport properties  in a wide range of
temperatures between the resistivity maximum $T_b$ and the phonon
frequency $\omega_0$, which is in principle experimentally accessible. 
The enhancement of the mobility 
 is signaled by a marked downturn from the Arrhenius behavior (see the right
hand side of Fig. 2), and takes place in the whole polaronic regime
$\lambda\gtrsim 1$ (as we shall see below, this behavior is quite
general and is not restricted to the adiabatic case).



%
\begin{figure}[htbp]
\centerline{\resizebox{8cm}{!}{\includegraphics{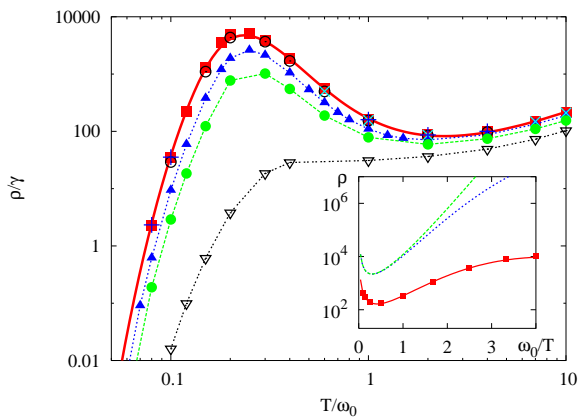}}}
\caption[]{Resistivity vs. temperature at $\alpha^2=10$.
The data points correspond respectively to $\gamma=$4 (+), 2 (squares), 1
(x), 0.5 (open circles), 0.2 (triangles), 0.15 (filled circles), 0.1
(open triangles) and were divided by $\gamma$ to evidence the
nonadiabatic scaling property (\ref{eq:scaling-anti}). Inset:
Arrhenius plot of $\rho$ at $\gamma=2$, compared with the perturbative
formulae --- see text.}
\label{fig:anti}
\end{figure}
%
In order to discuss the nonadiabatic case, where the polaron
formation is ruled by the parameter $\alpha^2$,  the
resistivity data are
illustrated in Fig. \ref{fig:anti} for different values of $\gamma$,
and fixed $\alpha^2=10$. 
We again recognize the three regimes of polaron
transport (coherent, activated, residual scattering).
The presence of such a ``peak-dip'' structure 
is therefore independent of the adiabaticity ratio $\gamma$ and {\it exists
whenever the carriers are of polaronic nature}. 
At large  $\gamma$, the resistivity obeys the following
scaling property
\begin{equation}
 \rho(T,\alpha^2,\gamma)=\gamma f(T/\omega_0,\alpha^2)  
\label{eq:scaling-anti}
\end{equation}
as shown in Fig.  \ref{fig:anti}.
Although the numerical integrals involved in eq. (\ref{eq:rho})
do not lead to an analytical expression for
$\rho(T)$, one can take advantage of the above scaling to
derive a tractable interpolation formula valid in the nonadiabatic regime.
To do so,
we choose to separate arbitrarily the  ``coherent'' part $\rho_C$ and 
the ``activated'' part $\rho_H$ by \textit{enforcing} Matthiessen's rule
$\rho^{-1}=\rho_C^{-1}+\rho_H^{-1}$. Introducing $y=T/\omega_0$,
the DMFT data are well described by
\begin{eqnarray}
  \label{eq:rho-anti-band}
  \rho_C(y)&=&A  \; 
  \gamma \ \alpha^4  y \ e^{ \alpha^2-1/y} \\
  \label{eq:rho-anti-act}
  \rho_H(y)&=&B \; \gamma \; y^{3/2}
  \exp \left\lbrack \Delta(y)/2 y \right\rbrack
\end{eqnarray}
with a temperature dependent activation gap 
\begin{equation}
   \label{eq:gap-anti}
  \Delta(y)=\alpha^2(1-\delta)
  \frac{\tanh{c/y}}{c/y} 
\end{equation}
and with $A=3.82$, $B=4.77$, $c=0.37$,
$\delta=0.26$.$^{17}$
Note that the form of the prefactors in
eqs. (\ref{eq:rho-anti-band})-(\ref{eq:rho-anti-act}) is constrained by the
scaling relation (\ref{eq:scaling-anti}).
The resulting curve for $\rho$ is plotted in Fig. \ref{fig:anti} for
$\alpha^2=10$ (full line). 
\footnote{The location 
of the resistivity maximum (for any $\alpha^2$ in the nonadiabatic regime) 
is in good agreement with the value given in ref.
\cite{LF} $T_b\simeq\omega_0/(2\log\alpha^2)$, while it  fundamentally 
contradicts Holstein's estimate of ref.\cite{Holstein}. The latter is based on 
the assumption  that the polaron bandwidth is reduced with
increasing temperature, which is a drawback of the perturbative
approach (any narrow feature in the
excitation spectrum, such as the polaron band, should rather 
get \textit{broadened} by thermal effects).}

Let us focus on the activated regime, as was done previously in the adiabatic
case. The problem  of polaron transport in this case is generally addressed  
from the  ``atomic'' limit $\gamma\to
\infty$\cite{Holstein,LF63}, treating the band parameter $D$ as a
perturbation. This yields the nonadiabatic textbook formula
\cite{Holstein,Mahan}
\begin{equation}
  \label{eq:anti-act}
  \rho(T)=B^\prime  \gamma^2 \alpha \; y^{3/2} \exp[\alpha^2/2y]
\end{equation}
valid at $T\gtrsim \omega_0$, with $B^\prime=(2^{7}/\pi)^{1/2}$  
[a generalization to  lower temperatures  was
given in ref. \cite{Holstein},  eq. (97)].
Since the effective expansion parameter which rules the perturbative treatment
is $\eta_2$ itself, eq. (\ref{eq:anti-act}) holds for  
$\eta_2\ll 1$, a condition opposite to  eq. (\ref{eq:condition}).
\cite{LF63}\footnote{Remarkably, at $\alpha^2=10$, the
  scaling property  (\ref{eq:scaling-anti}) holds for $\gamma\gtrsim
0.5$ which, taking   $T\approx \omega_0$,  corresponds  to $\eta_2\lesssim 1$.
Note, however,  that the low temperature coherent
regime is not ruled by eq. (\ref{eq:condition}),
 which relies on the analysis of incoherent hopping processes.} 

The inset of Fig. 3  shows an Arrhenius plot of the resistivity in 
the activated regime for $\alpha^2=10$ and $\gamma=2$. 
As in the adiabatic case, a marked downturn 
appears below  $T\approx \omega_0$,
indicating the onset of phonon quantum fluctuations. We infer that
this ubiquitous phenomenon is deeply related to the occurrence of a
resistivity maximum --- i.e.  of the very presence of small polarons --- 
as it takes place both in the adiabatic and
nonadiabatic regimes.

In the same inset, we have also drawn the resistivity given by 
eq. (\ref{eq:anti-act}) (dashed line), and
its low temperature generalization
(dotted line).
Compared to the DMFT results, we see that the perturbative formulae
wildly overestimate both the absolute value of the resistivity and the 
activation energy --- the slope of the curve --- 
\textit{within the activated regime}. Besides, a closer look at
eq. (\ref{eq:anti-act}) shows that it does not obey the
scaling formula (\ref{eq:scaling-anti}).
The disagreement is surprising, in view of the
fact that the chosen parameters ($\eta_2\approx 0.01$) 
lie well inside the range
of validity of the perturbative approach. 

The large discrepancy
comes from the narrow-band
character of the polaronic excitation spectrum.
In the limit $D\to 0$, 
the electron states are essentially independent on different sites. 
This, together with the fact that the phonons are assumed to be local and
dispersionless, prevents any transfer of energy between
sites (the spectral function is a distribution of delta peaks),  and
makes the transition probabilities singular.
Holstein healed the singularity  by introducing 
\textit{ad hoc} a sizeable phonon dispersion $\Delta\omega_{ph} \neq 0$, 
yielding eq. (\ref{eq:anti-act}).
However, especially in narrow band materials, 
the optical phonons often exhibit rather weak dispersions. 
Obtaining a finite result when
$\Delta\omega_{ph}\to 0$ requires to
treat the  \textit{electron} dispersion
(i.e. the finite bandwidth, $D\neq 0$) nonperturbatively. 
This can be achieved by the DMFT, as is testified by the finiteness of our results. 
%
\footnote{A simple calculation shows that, if $A(\nu)$ is a
collection of peaks of width $\Delta \omega$, the resistivity is
reduced by a factor  of $\Delta \omega/\omega_0$ relative to Holstein's result.
The observed scaling behavior
(\ref{eq:scaling-anti})
is recovered by noting that, if 
the broadening of the peaks is of 
electronic origin, then $\Delta \omega_{el} \propto D$.
In practical cases,
where both the electrons and the phonons disperse, it is likely that
the transport properties are determined by the larger of $\Delta
\omega_{el},\Delta \omega_{ph}$.}
%
%



In summary, we have applied the DMFT to study the transport 
properties of small polarons. The different behaviors expected by standard
polaron theory --- coherent, activated and residual scattering regime
---  are recovered within a unified  treatment, although 
notable deviations from the commonly accepted formulae are found.
First of all, a broad intermediate temperature regime emerges,
 regardless of the adiabaticity parameter $\gamma$, where 
the resistivity is still semiconducting-like, but is
strongly influenced by  phonon quantum fluctuations.
This regime, comprised  between the
resistivity maximum $T_b$ and the phonon frequency $\omega_0$, should be easily
detected experimentally as a downturn in the Arrhenius plots of the
resistivity. 
Secondly, in the nonadiabatic regime, the DMFT results obey a simple
scaling property, which is not compatible with the standard polaronic
formulae of Holstein. Accordingly, large quantitative discrepancies
arise in the predicted resistivity, which
could result in wrong estimates when extracting microscopic
parameters from the experiments.


Finally, we would like to point out that the adiabatic procedure which
was successfully
applied to study polaronic systems at finite density (half filling) within the
DMFT scheme,\cite{Millis} is not suitable at low densities,
where it
is not able to reproduce the correct activated behavior
(\ref{eq:adiab-act}). 
The reason is that the cooperative mechanism which leads to 
polaron localization at half
filling --- the presence of a bimodally distributed  random field
--- ceases to be effective at low electron densities, where the
phonons are \textit{not} renormalized. In the latter case, polaron
localization is a single-particle effect, the carriers being
self-trapped in their own local deformation. The relative
inportance of the two different mechanisms of polaron trapping can be
modulated by the density. The existence of an experimentally observable 
density crossover in polaronic
systems will be the subject of future investigation.

We acknowledge M. Capone for a critical reading of the manuscript.
%



\end{document}